\documentclass[journal,10pt]{IEEEtran}
\usepackage{amsmath, nccmath}
\usepackage{cases} 
\usepackage{amssymb}
\usepackage{cite}
\usepackage{url}
\usepackage{xcolor}
\usepackage{cite,graphicx}
\usepackage{subfigure}
\usepackage{citesort}
\usepackage{fancyhdr}
\usepackage{mdwmath}
\usepackage{mdwtab}
\usepackage{caption}
\usepackage{amsthm}
\usepackage{algorithmic}
\usepackage{algorithm}
\usepackage{accents}
\usepackage{stfloats}

\newtheorem{theorem}{Theorem}

\newtheorem{lemma}{Lemma}

\newtheorem{corollary}{Corollary}

\makeatletter
\def\ScaleIfNeeded{%
\ifdim\Gin@nat@width>\linewidth \linewidth \else \Gin@nat@width
\fi } \makeatother

\begin{document}

\title{On the Sum Rate and User Fairness of STAR-RIS Aided  Communications}
\author{
        Haochen~Li,~\IEEEmembership{Graduate Student Member,~IEEE,}
        Xidong~Mu,~\IEEEmembership{Member,~IEEE,}
        Yuanwei~Liu,~\IEEEmembership{Fellow,~IEEE,}       
        Yue~Chen,~\IEEEmembership{Senior Member,~IEEE,}
        Pan~Zhiwen,~\IEEEmembership{Member,~IEEE}

\thanks{Haochen~Li and Pan~Zhiwen are with National Mobile Communications Research Laboratory, Southeast University, China (email: lihaochen@seu.edu.cn, pzw@seu.edu.cn).}
\thanks{Xidong~Mu is with the Centre for Wireless Innovation, Queen's University Belfast, U.K. (e-mail: x.mu@qub.ac.uk).}
\thanks{Yuanwei Liu is with the Department of Electrical and Electronic Engineering, The University of Hong Kong, Hong Kong. (e-mail: yuanwei@hku.hk).}
\thanks{Yue~Chen is with the School of Electronic Engineering and Computer Science, Queen Mary University of London, U.K. (e-mail: yue.chen@qmul.ac.uk).}
}

\maketitle
\begin{abstract}
A simultaneously transmitting and reflecting reconfigurable intelligent surface (STAR-RIS) aided communication system is investigated. {A robust joint beamforming design problem under the imperfect channel state information (CSI) is formulated to maximize the weighted sum of the Jain's fairness index and the normalized system sum rate. To solve this non-convex problem, an alternating optimization (AO) algorithm is proposed, which leverages the S-Procedure, successive convex approximation (SCA), and semidefinite relaxation (SDR).  Simulation results demonstrate that with proposed algorithm: 1) various trade-offs between sum rate and user fairness can be achieved; 2) a larger trade-off region can be achieved by adopting STAR-RIS compared to conventional RIS; and 3) the performance degradation caused by imperfect CSI is less than $7\%$ with our proposed robust beamforming approach.}
\end{abstract}


\section{Introduction}

{To meet the growing data rate demands in next-generation communications, technologies such as multiple-input multiple-output (MIMO) and reconfigurable intelligent surfaces (RISs) have emerged as key enablers~\cite{9424177}. As traditional RISs are limited to half-space coverage by reflecting signals, simultaneously transmitting and reflecting RISs (STAR-RISs), introduced in~\cite{9570143}, overcome this limitation and provide full-space coverage. In STAR-RIS aided MIMO communication systems, a critical challenge is the design of the joint beamforming at the base station (BS) and STAR-RIS to achieve a satisfactory system data rate~\cite{10550177}.}

Maximizing the system sum rate is a widely used approach to address this issue. {In~\cite{10740607}, the authors proposed a fractional programming-based algorithm for joint beamforming design aiming at maximizing the sum rate in STAR-RIS aided MIMO systems. Moreover, the authors of~\cite{10439654} addressed the sum rate maximization problem in discrete and coupled phase-shift STAR-RIS aided communication systems. Furthermore, the authors of~\cite{10320337} investigated the sum rate maximization problem in STAR-RIS aided non-orthogonal multiple access networks.


 Note that wireless communication systems are typically operated with limited resources. The sum rate maximization only focuses on the overall system performance, and thus often overlooks the performance of individual users. With this drawback, the sum rate maximization approach could lead to fairness issues among users~\cite{8886335}. To address this problem, the authors of~\cite{10264820} proposed optimizing the max-min signal-to-noise ratio to guarantee user fairness in STAR-RIS aided MIMO systems. Additionally, the concept of max-min energy efficiency optimization was introduced in~\cite{10130543} to achieve fair energy efficiency design in STAR-RIS aided
MIMO systems. Moreover, the fairness-aware design approach has been extended to max-min communication and sensing performance optimization in STAR-RIS aided integrated sensing and communication beamforming design~\cite{10311519}. Although the max-min design can ensure fairness among users, the overall system performance may degrade since this approach forces the BS to provide equal quality of service to all users. In this case, the overall system performance is limited by the user with unfavorable channel conditions. The aforementioned works focus solely on either the sum rate or user fairness as the design criterion and lack a design approach that addresses the trade-off between the sum rate and user fairness.

Against this background, this work aims to balance the sum rate and user fairness for STAR-RIS aided communications, rather than prioritizing one over the other. To characterize the trade-off between sum rate and user fairness, we adopt a bi-criterion optimization framework that maximizes the weighted sum of the normalized sum rate and Jain's fairness index. Furthermore, we consider robust beamforming under imperfect channel state information (CSI), addressing the challenges of estimating STAR-RIS related channels due to its passive nature. To solve the resulting non-convex problem, we develop an alternating optimization (AO) algorithm that integrates the S-Procedure, successive convex approximation (SCA), and semi-definite relaxation (SDR). Simulation results demonstrate that: 1) various trade-offs between sum rate and user fairness can be achieved by adjusting the priority coefficient in the bi-criterion optimization objective; 2) the sum rate-fairness trade-off region is effectively expanded in STAR-RIS aided systems compared to conventional RIS aided systems; and 3) the performance degradation due to imperfect CSI is less than $7\%$ with the proposed robust joint beamforming design.}

\section{System Model}

Consider a STAR-RIS aided communication system, where a $M$-antenna BS communicates with $K$ single-antenna users with the help of an $N$-element STAR-RIS\footnote{{The energy splitting model is adopted because all elements in the STAR-RIS can simultaneously handle both transmission and reflection under this model, making it suitable for scenarios involving users on both sides of the STAR-RIS.}}. {The direct BS-user links are obstructed by obstacles, necessitating the use of the STAR-RIS to establish BS-STAR-user connections in the signal blind spots.} Users positioned on both sides of the STAR-RIS are designated as T-users and R-users, respectively.{ Denote the sets of users and STAR elements as $\mathcal{K}$ and $\mathcal{N}$, respectively. Denote the sets of T-users and R-users as $\mathcal{K}_t=\left\{k\in\mathbb{Z}|1\le k\le \bar{K}\right\}$ and $\mathcal{K}_r=\left\{k\in\mathbb{Z}|\bar{K}+1\le k\le K\right\}$, respectively, where $\bar{K}$ is the number of T-users.}

Let $\theta_{n}^i$ and $\beta_{n}^i$ represent the phase shift and amplitude coefficients, respectively, where $i \in \{t, r\}$ and~$n \in \mathcal{N}$. The STAR-RIS phase shift matrices are  
\begin{equation}
    \mathbf{\Phi}_i=\mathrm{diag}({\beta_{1}^i}e^{j\theta_{1}^i}, {\beta_{2}^i}e^{j\theta_{2}^i}, \cdots, {\beta_{N}^i}e^{j\theta_{N}^i}), i\in\{t,r\}.
\end{equation} 
It is assumed that amplitude coefficients $\beta_{n}^i$ can vary continuously within the interval $\left[0, 1\right]$ and satisfies the energy conservation constraint $\left(\beta_{n}^t\right)^2 + \left(\beta_{n}^r\right)^2 = 1$.

The signal arrived at T users $ k \in \mathcal{K}_t$  or R users  $ k \in \mathcal{K}_r$ can be expressed as:
\begin{equation}
\begin{aligned}
y_{k}=\mathbf{g}_{k}^H\mathbf{\Phi}_{i}\mathbf{G}\mathbf{w}_{k}s_{k}+\mathbf{g}_{k}^H\mathbf{\Phi}_{i}\mathbf{G}\sum\nolimits_{j\in \mathcal{K},j\ne k}\mathbf{w}_{j}s_{j}+n_k,
\end{aligned}
\end{equation}
where $\mathbf{w}_{k}$ and $s_{k}$ represent the beamforming vector of the BS and signal for user $k \in \mathcal{K}_i, \forall i \in \{t,r\}$, respectively. The matrix $\mathbf{G} \in \mathbb{C}^{N \times M}$ denotes the channel from the BS to the STAR-RIS, while $\mathbf{g}_k \in \mathbb{C}^{N \times 1}$ represents the channel from the STAR-RIS to user $k$. The noise at user $k$, denoted as $n_k$, follows a complex normal distribution $\mathcal{CN}(0, \sigma^2_k)$. The communication capacity of user~$k$  can be given as

\begin{equation}
\begin{aligned}
R_{k}=\log \Bigg(1+\frac{\left|\mathbf{v}_{i}^H\mathbf{H}_k\mathbf{w}_{k}\right|^{2}}{\sum\nolimits_{j\in \mathcal{K},j\ne k}\left|\mathbf{v}_{i}^H\mathbf{H}_k\mathbf{w}_{j}\right|^{2}+\sigma^{2}_k}\Bigg),
\end{aligned}
\end{equation}
where $\mathbf{v}_i=\text{diag}\left(\mathbf{\Phi}_i^H\right)$, $\forall i\in\{t,r\}$. $\mathbf{H}_k=\text{diag}\left(\mathbf{g}_k^H\right)\mathbf{G}$, $k\in\mathcal{K}$, is the cascaded channel of user $k$. {The cascaded channel is challenging to estimate due to the passive nature of the STAR-RIS. Hence, it is assumed that the CSI of cascaded channels is imperfect with the bounded error model~\cite{9180053}:
\begin{equation}
\begin{aligned}
\mathbf{H}_k=\bar{\mathbf{H}}_k+\Delta\mathbf{H}_k,\mathcal{E}_k=\{\mathbf{H}_k | \|\Delta\mathbf{H}_k\|_F\le \varepsilon_k\},\forall k,
\end{aligned}
\end{equation}
where $\varepsilon_k$ is the radii of uncertainty region $\mathcal{E}_k$.} The normalized sum rate and Jain's fairness index can be given by

\begin{equation}\label{equation:sum_rate}
\begin{aligned}
R\left(\mathbf{w},\mathbf{v}\right)=\frac{\sum\nolimits_{k\in \mathcal{K}}R_k}{R_\text{max}},
\end{aligned}
\end{equation}
\begin{equation}\label{equation:jain}
\begin{aligned}
J\left(\mathbf{w},\mathbf{v}\right)=\frac{1}{K}\frac{\big(\sum\nolimits_{k\in \mathcal{K}}R_k\big)^2}{\sum\nolimits_{k\in \mathcal{K}}R_k^2},
\end{aligned}
\end{equation}
where $\mathbf{w}\buildrel \Delta \over=\{\mathbf{w}_1,\mathbf{w}_2,\cdots,\mathbf{w}_K\}$ and $\mathbf{v}\buildrel \Delta \over=\{\mathbf{v}_t,\mathbf{v}_r\}$. $R_\text{max}$ is the system's maximum sum rate, which can be obtained by solving the conventional sum rate maximization problem.

{To characterize the trade-off between sum rate and user fairness, we use a bi-criterion optimization framework that maximizes the weighted sum of the normalized sum rate and Jain's fairness index, which results the below problem:
\begin{subequations}\label{problem:sum_rate_fairness}
    \begin{align} 
    \label{problem:sum_rate_fairness_obj}       
        \max_{\mathbf{w}, \mathbf{v}} \quad &  \alpha R\left(\mathbf{w},\mathbf{v}\right) + (1-\alpha) J\left(\mathbf{w},\mathbf{v}\right) \\
        \label{constraint:STAR_1} 
        \mathrm{s.t.} \quad & \left({\rho}_n^t\right)^2+\left({\rho}_n^r\right)^2=1, \forall n,\\
        \label{constraint:STAR_2} 
        & {\rho}_n^k\in [0,1], \forall n, \forall k,\\
        \label{constraint:STAR_3}       
        & {\theta}_n^t\in [0,2\pi), \forall n, \forall k,\\
        \label{constraint:power}
        & \sum\nolimits_{k\in \mathcal{K}}\|\mathbf{w}_k\|^2  \le P,
    \end{align}
\end{subequations}
where} {$0 < \alpha \le 1$ denotes the priority coefficient that balances the significance of the sum rate against Jain's fairness index. The motivation behind the objective function  is to strike a balance between maximizing system performance and ensuring fairness among users.} The constraints specified in~\eqref{constraint:STAR_1}-\eqref{constraint:STAR_3} are related to the limits of the STAR-RIS coefficients. Constraint~\eqref{constraint:power} specifies the power limitation at the BS.


{\section{Proposed Optimization Method}
To address it, we first reformulate~problem~\eqref{problem:sum_rate_fairness} as follows:
{\begin{subequations}\label{problem:sum_rate_fairness_slack}
    \begin{align}        
        \max_{\mathbf{w}, \mathbf{v}, a_R, a_J}   \quad &  a_R+ a_J \\
        \label{constraint:a_r}
        \mathrm{s.t.} \quad & \alpha R(\mathbf{w},\mathbf{\Phi})  \ge a_R,\\
        \label{constraint:a_j}
        &(1-\alpha) J(\mathbf{w},\mathbf{\Phi}) \ge a_J,\\
        & \eqref{constraint:STAR_1}-\eqref{constraint:power},
    \end{align}
\end{subequations}
{where $a_R$ represents the relaxed normalized sum rate, while $a_J$ corresponds to the relaxed Jain’s fairness index. These variables enable us to decouple the bi-criterion optimization problem, thereby simplifying its reformulation.}  Introducing slack variables $c_k, k\in\mathcal{K}$, and substituting~\eqref{equation:sum_rate} and~\eqref{equation:jain} into constraints~\eqref{constraint:a_r} and~\eqref{constraint:a_j}, we have
\begin{subnumcases}{\eqref{constraint:a_r},\eqref{constraint:a_j}  \Leftrightarrow  }
   {\sum\nolimits_{k\in \mathcal{K}}{c_k} \ge \frac{{{R_{{\rm{max}}}}{a_R}}}{\alpha }}, \label{slack_R}
   \\
    {\frac{{{{\Big(\sum\nolimits_{k\in \mathcal{K}}{c_k}\Big)}^2}}}{\sum\nolimits_{k\in \mathcal{K}}{c_k^2}} \ge \frac{{K{a_J}}}{{1 - \alpha }}}, \label{slack_J} \\
    {{R_k} \ge {c_k},\forall k \in \mathcal{K} }.\label{c}
\end{subnumcases}}

\noindent It can be observed that the constraints~\eqref{slack_J} and~\eqref{c} are non-convex due to the fractional form in the Jain's fairness index and the achievable data rate of users. Introducing slack variables $\tau$ and $\mu$, \eqref{slack_J} can be reformulated as follows:
\begin{subnumcases}{\eqref{slack_J}  \Leftrightarrow  }
    \Big(\sum\nolimits_{k\in \mathcal{K}}c_k\Big)^2 \ge \mu^2\tau^2 \Leftrightarrow \sum\nolimits_{k\in \mathcal{K}}c_k \ge \mu\tau , \label{slack_J1}
    \\
    \sum\nolimits_{k\in \mathcal{K}}c_k^2 \le \tau^2 \buildrel (a) \over\Leftrightarrow \|[c_1\ c_2\ \cdots \  c_K]^T\| \le \tau, \label{slack_J2} \\
    \mu^2 \ge \frac{Ka_J}{1-\alpha},\label{slack_J3}
\end{subnumcases}
where $(a)$ shows that constraint~\eqref{slack_J2} is convex as it can be rewritten in the form of a second-order cone.
 {Introducing slack variables $\eta_k$ and $\lambda_k$,~\eqref{c} can be reformulated as follows for $\forall \mathbf{H}_k\in\mathcal{E}_k$,
\begin{subnumcases}{\!\eqref{c} \! \Leftrightarrow \! }
    \text{vec}(\mathbf{H}_k)^H(\mathbf{W}_{{k}}^T\otimes\mathbf{V}_i)\text{vec}(\mathbf{H}_k)^H\ge\eta_k\lambda_k,  \label{slack_c1}
    \\
    \text{vec}(\mathbf{H}_k)^H(\mathbf{W}_{\bar{k}}^T\otimes\mathbf{V}_i)\text{vec}(\mathbf{H}_k)^H+\sigma_k^2 \le \lambda_k,  \label{slack_c2} \\
    \eta_k \ge 2^{c_k}-1,\label{slack_c3}
\end{subnumcases}
where $\mathbf{W}_k=\mathbf{w}_k\mathbf{w}_k^H$, $\mathbf{W}_{\bar{k}}=\sum\nolimits_{j\ne k}\mathbf{W}_j$, and $\mathbf{V}_k=\mathbf{v}_i\mathbf{v}_i^H$. Due to the uncertainty of $\mathbf{H}_k$, there are actually an infinite number of such constraints~\eqref{slack_c1} and~\eqref{slack_c2}. The S-Procedure~\cite[Appendix~B.2]{boyd2004convex} is used to deal with this challenge by transforming the constraints~\eqref{slack_c1} and~\eqref{slack_c2}  into following linear matrix inequality constraints for $\forall k \in \mathcal{K}_i, \forall i \in \{t,r\}$
\begin{equation}\label{LMI1}
       \left[ {\begin{array}{*{20}{c}}
{\mathbf{A}_k+\delta_k\mathbf{I }}&{\mathbf{A}_k\text{vec}(\bar{\mathbf{H}}_k)}\\
{\text{vec}(\bar{\mathbf{H}}_k)^H\mathbf{A}_k}&{\text{vec}(\bar{\mathbf{H}}_k)^H\mathbf{A}_k\text{vec}(\bar{\mathbf{H}}_k)+a_k}
\end{array}} \right]\succeq 0,
\end{equation}
\begin{equation}\label{LMI2}
       \left[ {\begin{array}{*{20}{c}}
{\mathbf{B}_k+\kappa_k\mathbf{I}_{MN}}&{\mathbf{B}_k\text{vec}(\bar{\mathbf{H}}_k)}\\
{\text{vec}(\bar{\mathbf{H}}_k)^H\mathbf{B}_k}&{\text{vec}(\bar{\mathbf{H}}_k)^H\mathbf{B}_k\text{vec}(\bar{\mathbf{H}}_k)+b_k}
\end{array}} \right]\succeq 0,
\end{equation}
where $\delta_k$, $\kappa_k\ge 0$ are slack variables introduced by the S-Procedure, $\mathbf{A}_k=\mathbf{W}_{{k}}^T\otimes\mathbf{V}_i$, $\mathbf{B}_k=-\mathbf{W}_{\bar{k}}^T\otimes\mathbf{V}_i$, $a_k=-\eta_k\lambda_k-\delta_k\varepsilon_k^2$, and $b_k=-\sigma_k^2+\lambda_k-\kappa_k\varepsilon_k^2$.
Constraints~\eqref{constraint:STAR_1}-\eqref{constraint:power} can be equivalent represented as follows: 

\begin{subnumcases}{\begin{array}{c}
\eqref{constraint:STAR_1}$-$\eqref{constraint:power}
\end{array}\Leftrightarrow}
    {\left[\mathbf{V}_t\right]_{nn}+\left[\mathbf{V}_r\right]_{nn}=1, \left[\mathbf{V}_i\right]_{nn}\in \left[0,1\right]},\label{v}\\
    {\sum\nolimits_{k \in \mathcal{K}}\text{tr}\left(\mathbf{W}_k\right)\le P},\label{w}\\
    {\mathbf{W}_k \succeq 0, \mathbf{V}_i \succeq 0, \forall k \in \mathcal{K}},\label{semi_wv}\\
    { \text{rank}\left(\mathbf{W}_k\right)=1, \text{rank}\left(\mathbf{V}_i\right)=1},\label{rank_wv}
\end{subnumcases}
where notation $\left[\cdot\right]_{nn}$ extracts the $n$-th element on the main diagonal of the matrix. Substituting reformulations in~(8)-(13) into problem~\eqref{problem:sum_rate_fairness_slack} and carrying out the SDR by removing constraint~\eqref{rank_wv}, following optimization problem can be obtained
\begin{subequations}\label{problem:sum_rate_fairness_slack_all}
    \begin{align}   
        \label{obj:sum_rate_fairness_slack_all}   
        \!\!\!\max_{\mathbf{W},\mathbf{V},\mathbf{\chi}} \ &  a_R+ a_J \\
        \mathrm{s.t.} \ & \eqref{slack_R}, \eqref{slack_J1}-\eqref{slack_J3}, \eqref{slack_c3}, \eqref{LMI1}, \eqref{LMI2}, \eqref{v}-\eqref{semi_wv},
    \end{align}
\end{subequations}
where $\mathbf{\chi} \buildrel \Delta \over= \left\{a_R, a_J, \tau, \mu, c_k, \eta_k, \lambda_k, \delta_k, \kappa_k, \forall k \in \mathcal{K}\right\}$ denotes the set containing all slack variables. The AO method is adopted to address problem~\eqref{problem:sum_rate_fairness_slack_all} with variables in two groups, i.e., ${W} \buildrel \Delta \over= \left\{\mathbf{W}_1, \mathbf{W}_2, \cdots, \mathbf{W}_K\right\}$ and $\mathbf{V} \buildrel \Delta \over= \left\{\mathbf{V}_t, \mathbf{V}_r\right\}$.

\subsection{Optimizing $\mathbf{W}$ with Fixed $\mathbf{V}$}
Given STAR-RIS transmission and reflection matrices, the subproblem with respect to BS beamforming can be expressed as follows: 
\begin{subequations}\label{problem:sum_rate_fairness_slack_all_w}
    \begin{align}   
        \label{obj:sum_rate_fairness_slack_all_w}   
        \!\!\!\max_{\mathbf{W},\mathbf{\chi}} \ &  a_R+ a_J \\
        \mathrm{s.t.} \ & \eqref{slack_R}, \eqref{slack_J1}-\eqref{slack_J3}, \eqref{slack_c3}, \eqref{LMI1}, \eqref{LMI2}, \eqref{w}, \eqref{semi_wv}.
    \end{align}
\end{subequations}
Problem~\eqref{problem:sum_rate_fairness_slack_all_w} is non-convex due to the presence of non-convex constraints~\eqref{slack_J1},~\eqref{slack_J3}, and~\eqref{LMI1}. The SCA is then employed to tackle this difficulty with the following approximation
\begin{equation}\label{ab_le}
\begin{aligned}
   &{f}_\text{lb}\left(a,b,a_0,b_0\right)=\frac{1}{2}(a_0+b_0)^2-\frac{1}{2}a^2-\frac{1}{2}b^2\\
   &+\left(a_0+b_0\right)\left(a-a_0+b-b_0\right) \le ab \le \frac{1}{2}(a+b)^2\\
   &+\frac{1}{2}a_0^2-\frac{1}{2}a_0a+\frac{1}{2}b_0^2-\frac{1}{2}b_0b={f}_\text{ub}\left(a,b,a_0,b_0\right).
\end{aligned}
\end{equation}
where  ${a}$, ${b}$, ${a}_0$, and $b_0$ can be any complex scalars. ${f}_\text{lb}(\cdot)$ and ${f}_\text{ub}(\cdot)$ denote the lower bound concave function and the upper bound convex function, respectively. The proof for aforementioned inequality is given as follows. For any complex scalars $a$ and $b$, it holds that \begin{equation}\label{AB}
    ab=\frac{1}{2}(a+b)^2-\Big(\frac{1}{2}a^2+\frac{1}{2}b^2\Big).
\end{equation}
The right-hand side of~\eqref{AB} is in the form of a difference of convex (DC) functions. The global lower bounds of the convex terms in~\eqref{AB} can be obtained using the first-order Taylor expansion at the given point $\left\{a_0, b_0\right\}$, i.e.,
\begin{equation}\label{AB_1}
\begin{aligned}
    \!\!\!\!(a+b)^2 \ge (a_0+b_0)^2+2\left(a_0+b_0\right)\left(a-a_0+b-b_0\right),
\end{aligned}
\end{equation} 
\begin{equation}\label{AB_2}
\begin{aligned}
     x^2 \ge x_0^2+2x_0\left(x-x_0\right)=2x_0x-x_0^2,
\end{aligned}
\end{equation} 
where $\left\{x, x_0\right\}$ could be $\left\{a, a_0\right\}$ or $\left\{b, b_0\right\}$. Substituting~\eqref{AB_1} into~\eqref{AB}, the upper bound in~\eqref{ab_le} can be obtained. Substituting~\eqref{AB_2} into~\eqref{AB}, the lower bound in~\eqref{ab_le} can be obtained.



Adopting aforementioned Theorem at given point $\bar{\mathbf{\chi}}^{(l)}=\{\mu^{(l)},\tau^{(l)},\eta_k^{(l)},\lambda_k^{(l)}\}$ in the $l$-th SCA iteration, non-convex constraints~\eqref{slack_J1},~\eqref{slack_J3}, and~\eqref{LMI1} can be approximated as
\begin{equation}\label{slack_J1_appr}
    \sum\limits_{k\in \mathcal{K}}c_k \ge {f}_\text{ub}\left(\mu,\tau,\mu^{(l)},\tau^{(l)}\right),
\end{equation}
\begin{equation}\label{slack_J3_appr}
      {f}_\text{lb}\left(\mu,\mu,\mu^{(l)},\mu^{(l)}\right) \ge \frac{Ka_J}{1-\alpha},
\end{equation}
\begin{equation}\label{LMI1_appr}
       \left[ {\begin{array}{*{20}{c}}
{\!\!\mathbf{A}_k+\delta_k\mathbf{I }}&{\mathbf{A}_k\text{vec}(\bar{\mathbf{H}}_k)\!\!}\\
{\!\!\text{vec}(\bar{\mathbf{H}}_k)^H\mathbf{A}_k}&{\text{vec}(\bar{\mathbf{H}}_k)^H\mathbf{A}_k\text{vec}(\bar{\mathbf{H}}_k)+a_k^{(l)}\!\!}
\end{array}} \right]\succeq 0,
\end{equation}
where $a_k^{(l)}=-{f}_\text{ub}\left(\eta_k,\lambda_k,\eta_k^{(l)},\lambda_k^{(l)}\right)-\delta_k\varepsilon_k^2$. Problem~\eqref{problem:sum_rate_fairness_slack_all} can be approximated~as
\begin{subequations}\label{problem:sum_rate_fairness_slack_all_app_w}
    \begin{align}   
        \label{obj:sum_rate_fairness_slack_all_app_w}   
        \max_{\mathbf{W},\mathbf{\chi}} \quad &  a_R+ a_J \\
        \mathrm{s.t.} \quad  & \eqref{slack_R}, \eqref{slack_J2}, \eqref{slack_c3}, \eqref{LMI2}, \eqref{w}, \eqref{semi_wv}, \eqref{slack_J1_appr}-\eqref{LMI1_appr}.
    \end{align}
\end{subequations}
The approximated problem~\eqref{problem:sum_rate_fairness_slack_all_app_w} is convex and can therefore be optimally solved using standard convex solvers with computational complexity of $\mathcal{O}(KM^{3.5})$~\cite{5447068}. 

\subsection{Optimizing $\mathbf{V}$ with Fixed $\mathbf{W}$}
Given BS beamforming matrices, the subproblem with respect to STAR-RIS beamforming can be expressed as follows: 
\begin{subequations}\label{problem:sum_rate_fairness_slack_all_v}
    \begin{align}   
        \label{obj:sum_rate_fairness_slack_all_v}   
        \!\!\!\max_{\mathbf{V},\mathbf{\chi}} \ &  a_R+ a_J \\
        \mathrm{s.t.} \ & \eqref{slack_R}, \eqref{slack_J1}-\eqref{slack_J3}, \eqref{slack_c3}, \eqref{LMI1}, \eqref{LMI2}, \eqref{v}, \eqref{semi_wv}.
    \end{align}
\end{subequations}
The non-convexity of problem~\eqref{problem:sum_rate_fairness_slack_all_v}   also results from constraints~\eqref{slack_J1},~\eqref{slack_J3}, and~\eqref{LMI1}. Thus, the SCA method same  to the former subsection is adopted to approximate these constraints at given point $\hat{\mathbf{\chi}}^{(l)}=\{\hat{\mu}^{(l)},\hat{\tau}^{(l)},\hat{\eta}_k^{(l)},\hat{\lambda}_k^{(l)}\}$ in the $l$-th SCA iteration. The resulted approximated problem is convex and can therefore be optimally solved using standard convex solvers with computational complexity of $\mathcal{O}(2N^{3.5})$~\cite{5447068}. 

\subsection{Overall Algorithm}
{\begin{algorithm}[tbp]
\caption{The AO algorithm for solving problem~\eqref{problem:sum_rate_fairness_slack_all}.}\label{algorithm1}
\begin{algorithmic}[1]
{\STATE {Initialize $\mathbf{W}^{(0)}$, $\mathbf{V}^{(0)}$, $\bar{\mathbf{\chi}}^{(0)}$, and $\epsilon_{\text{SCA}}$, and set $l=0$.}\\
\STATE {\bf repeat: }\\
\STATE \quad Given $\mathbf{V}^{(l)}$ and $\bar{\mathbf{\chi}}^{(l)}$, obtain $\mathbf{W}^{(l+1)}$ and $\hat{\mathbf{\chi}}^{(l+1)}$.\\
\STATE \quad Given $\mathbf{W}^{(l+1)}$ and $\hat{\mathbf{\chi}}^{(l+1)}$, obtain $\mathbf{\Phi}^{(l+1)}$ and ${\bar{\mathbf{\chi}}^{(l+1)}}$.\\
\STATE \quad Set $l =l+1$.\\
\STATE {\bf until} the increase in~\eqref{obj:sum_rate_fairness_slack_all} is below~$\epsilon_{\text{SCA}}$.}
\end{algorithmic}
\end{algorithm}}
Building on the previously obtained results, we propose an iterative algorithm to address problem~\eqref{problem:sum_rate_fairness_slack_all}. In each iteration, this algorithm solves the convex approximation of problem~\eqref{problem:sum_rate_fairness_slack_all_w} and problem~\eqref{problem:sum_rate_fairness_slack_all_v} sequentially. The algorithm's details are outlined in \textbf{Algorithm~\ref{algorithm1}}. {\text{Algorithm~\ref{algorithm1}} can converge to at least a stationary point of problem~\eqref{problem:sum_rate_fairness_slack_all} as the objective value of~\eqref{problem:sum_rate_fairness_slack_all} is bounded above and non-decreasing in each iteration. The computational complexity of {Algorithm~\ref{algorithm1}} is $\mathcal{O}(I(KM^{3.5}+2N^{3.5}))$, where $I$ is the  number of iterations. }}

Then, we discuss the tightness of the SDR with respect to $\mathbf{W}$ and $\mathbf{V}$.  For the tightness of the SDR with respect to $\mathbf{W}$, we have the following theorem.
\begin{theorem}\label{Theorem3} \emph{Given any $P>0$, the optimal solution $\mathbf{W}^* \buildrel \Delta \over= \left\{\mathbf{W}_1^*, \mathbf{W}_2^*, \cdots, \mathbf{W}_K^*\right\}$ that satisfies $\text{rank}\left(\mathbf{W}_k^*\right)\le1, \forall k \in \mathcal{K}$ can always be obtained.}
\begin{proof}
{Please refer to Appendix A.}
\end{proof}
\end{theorem}
In contrast, the optimal solution $\mathbf{V}^* \buildrel \Delta \over= \left\{\mathbf{V}_t^*, \mathbf{V}_r^*\right\}$ typically has general rank. To tackle this difficulty, the Gaussian randomization can be adopted to acquire the STAR-RIS coefficient vectors $\hat{\mathbf{v}}_t$ and $\hat{\mathbf{v}}_r$ from $\mathbf{V}^*$. Considering that the STAR-RIS coefficients need to satisfy the constraint~\eqref{constraint:STAR_1}, the STAR-RIS coefficient vectors are further normalized as follows:
\begin{equation}
  [\mathbf{v}_i]_n = \frac{[\hat{\mathbf{v}}_i]_n}{\sqrt{\left|[\hat{\mathbf{v}}_t]_n\right|^2+\left|[\hat{\mathbf{v}}_r]_n\right|^2}}, \forall i \in\{t,r\}, \forall n \in\mathcal{N}.
\end{equation}

\section{Simulation Results} \label{sec:results}
\begin{figure} [htbp]
\centering \vspace{-0.1cm}
\includegraphics[width=0.35\textwidth]{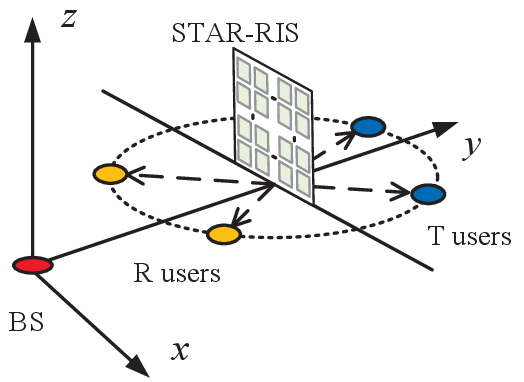}
 \caption{The simulation setup.
  }\vspace{-0.1cm}
 \label{setup}
\end{figure}

{The simulation setup is shown in Fig.~\ref{setup}, where the BS is located at coordinates $\left({0, 0, 0}\right)$ meters, and the STAR-RIS is positioned at $\left({0, 30, 0}\right)$ meters. All users are arranged in a concentric ring centered at the STAR-RIS, with an inner radius of $3$ meters and an outer radius of $8$ meters.} The channel path-loss coefficient is modeled by $\beta=C_{0}(\frac{d}{D_{0}})^{-\alpha}$, where $C_{0}=-30$ dB is the path loss at reference distance $D_{0}=1$ meter.  Here, $d$ represents the distance, and $\alpha=2.2$ is the path loss exponent. Small-scale fading between the BS and the STAR-RIS, as well as between the STAR-RIS and the users, is modeled using Rician fading with a Rician factor of $3$ dB. {The uncertain degree of the estimation
error for cascaded channel between the BS and user $k$ is given by $\rho_k=\frac{\varepsilon_k^2}{\|\bar{\mathbf{H}}_k\|_F^2}, \forall k$. Additional system parameters are as follows, unless stated otherwise: $P=30$ dBm (BS power budget), $\sigma^{2}=-90$ dBm (noise power at users), $\rho_k=\rho,\forall k$, $N=40$ (number of RIS elements), $M=8$ (number of BS antennas), $K=4$ (number of users), and $\bar{K}=2$ (number of T-users).
}

\begin{figure} [!t]
\centering
\includegraphics[width=0.33\textwidth]{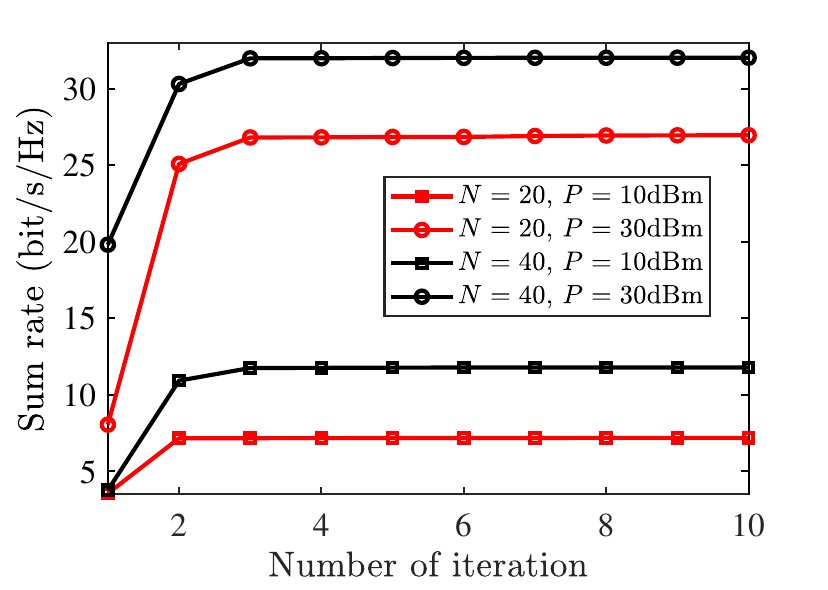} 
 \caption{{Convergence behavior of {Algorithm~\ref{algorithm1}}}.
  }\vspace{-0.1cm}
 \label{convergence}
\end{figure}

\begin{figure} [!t]
\centering
\includegraphics[width=0.33\textwidth]{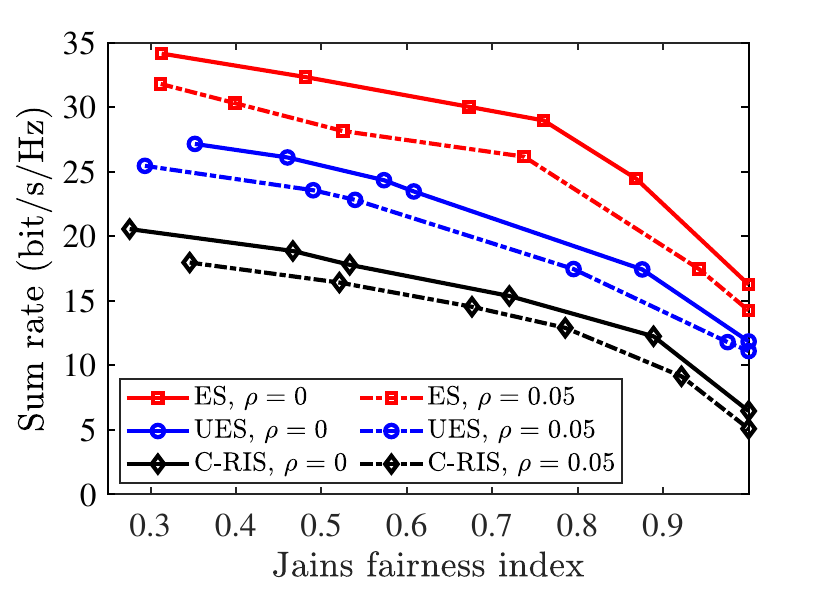}
 \caption{{The rate-fairness trade-off regions of different schemes.}
  }\vspace{-0.1cm}
 \label{Pareto}
\end{figure}

\begin{figure} [!t]
\centering
\includegraphics[width=0.33\textwidth]{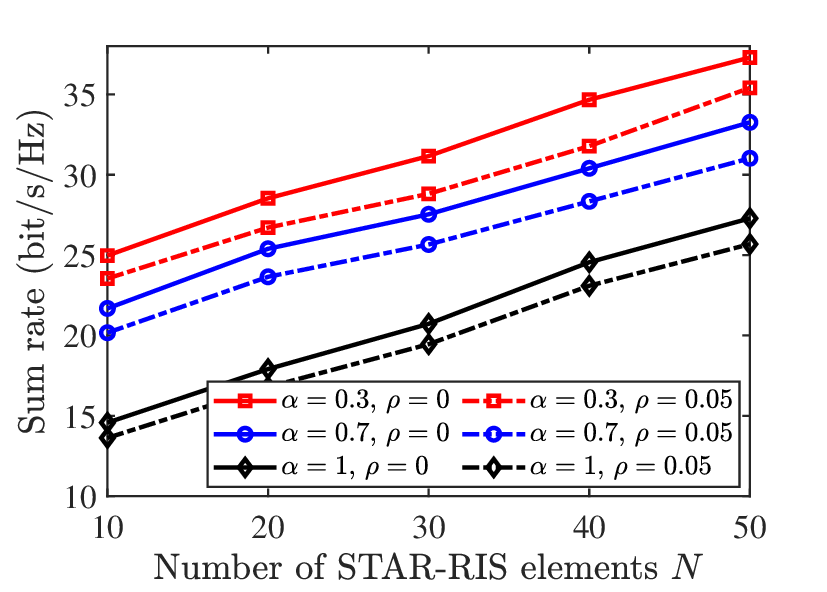}
 \caption{{The sum rate against number of STAR-RIS elements.}
  }\vspace{-0.1cm}
 \label{Min}
\end{figure}

The baseline schemes \textbf{uniform energy splitting STAR-RIS} and \textbf{Conventional RIS} used in~\cite{9570143} are utilized for comparison with our proposed approach to verify the effectiveness of the algorithm. {In the simulation figures, the legends "ES," "UES," and "C-RIS" represent the application of the proposed algorithm to communication systems with energy-splitting STAR-RIS, uniform energy-splitting STAR-RIS, and conventional RIS, respectively.}

{We begin by examining the convergence behavior of the proposed algorithm with $\rho= 0.05$. Fig.~\ref{convergence} illustrates both the convergence speed and the achievable sum rate for the proposed algorithm. For all simulation configurations, the proposed algorithms converge within $4$ iterations. 
}


In Fig.~\ref{Pareto}, we investigate the trade-off between the system sum rate and user fairness by illustrating {their rate-fairness trade-off region.  Each point in Fig.~\ref{Pareto} stands for an achievable trade-off between sum rate and user fairness.  By adjusting the priority coefficient in the bi-criterion optimization objective of problem~\eqref{problem:sum_rate_fairness}, various sum rate and user fairness trade-offs can be obtained. Besides, it can be observed that the STAR-RIS aided communication system effectively expands the rate-fairness trade-off region. This is due to the fact that the STAR-RIS provides additional degrees of freedom for enhancing the desired signal and mitigating interference. Furthermore, the performance degradation caused by imperfect CSI is less than $7\%$ in the proposed STAR-RIS aided communication system, demonstrating the effectiveness of the proposed robust beamforming algorithm.}

{In Fig.~\ref{Min}, we illustrate the  user sum rate versus the number of RIS elements $N$ under different priority coefficients~$\alpha$. It is evident that, by decreasing the priority coefficient, the priority of user fairness is improved during the joint active and passive design, resulting in a diminished system sum rate. This observation aligns with the results from Fig.\ref{Pareto}. }

{\section{Conclusions}
{The robust joint beamforming design for STAR-RIS aided communication systems was investigated, with the goal of achieving a trade-off between sum rate and user fairness under imperfect CSI. The solution to the resulting non-convex problem was obtained through the developed AO algorithm with the S-Procedure, SCA, and SDR. Simulation results demonstrated that the proposed robust beamforming algorithm can achieve effective beamforming design with imperfect CSI and strike a balance between the system sum rate and user fairness.}}


\section*{{Appendix~A: Proof of Theorem~\ref{Theorem3}}} \label{Appendix:C}
\renewcommand{\theequation}{A.\arabic{equation}}
\setcounter{equation}{0}
{Strong duality holds for problem~\eqref{problem:sum_rate_fairness_slack_all_app_w} since it is convex and satisfies the Slater's condition. The terms with respect to $\mathbf{W}$ in Lagrangian function of~\eqref{problem:sum_rate_fairness_slack_all_app_w}  is $\mathcal{L}_k=\sum\nolimits_{k}\mathcal{L}_k$, where
\begin{equation}
\begin{aligned}
   \!\! \!\!\mathcal{L}_k=\Gamma\text{tr}(\mathbf{W}_k)-\text{tr}(\mathbf{Y}_k\!\mathbf{W}_k)-\text{tr}(\bar{\mathbf{Y}}_k\!\bar{\mathbf{X}}_k)-\text{tr}(\hat{\mathbf{Y}}_k\!\hat{\mathbf{X}}_k),
\end{aligned}
\end{equation}
wherein $\bar{\mathbf{X}}_k$ and $\hat{\mathbf{X}}_k$ stand for the LMIs in constraints~\eqref{LMI2} and~\eqref{LMI1_appr}, respectively  $\Gamma$ is the non-negative scalar Lagrange multiplier associated with constraint~\eqref{w}.  $\mathbf{Y}_k$, $\bar{\mathbf{Y}}_k$, and $\hat{\mathbf{Y}}_k$  are semi-definite matrices Lagrange multiplier associated with constraints~\eqref{semi_wv},~\eqref{LMI2}, and~\eqref{LMI1_appr}, respectively. The optimal solution $\mathbf{W}^*$ satisfies the  Karush-Kuhn-Tucker (KKT) condition. Thus, we have
\begin{equation}\label{C}
     \mathbf{Y}_k^*\succeq 0, \bar{\mathbf{Y}}_k^*\succeq 0, \hat{\mathbf{Y}}_k^*\succeq 0, \Gamma^* \ge 0,
\end{equation}
\begin{equation}
    \mathbf{Y}_k^*\mathbf{W}_k^* =\mathbf{0},
\end{equation}
\begin{equation}
    \nabla_{\mathbf{W}_k^*}\mathcal{L} =\mathbf{0} \Leftrightarrow \mathbf{Y}_k^*=\Gamma^*\mathbf{I}_M-\mathbf{C}_k,
\end{equation}
\begin{equation}
\begin{aligned}
    \mathbf{C}_k=&\sum_{i=1}^{N}\left[\mathbf{U}_k^H\hat{\mathbf{Y}}_k^*\mathbf{U}_k\right]_{(i-1)M:iM,(i-1)M:iM}\\
    &-\sum_{j\ne k}^{N}\sum_{i=1}^{N}\left[\mathbf{U}_j^H\bar{\mathbf{Y}}_j^*\mathbf{U}_j\right]_{(i-1)M:iM,(i-1)M:iM},
\end{aligned}
\end{equation}
where $\Gamma^*$, $\mathbf{Y}_k^*$,  $\bar{\mathbf{Y}}_k^*$, and $\hat{\mathbf{Y}}_k^*$  are optimal Lagrange multipliers. $\mathbf{U}_k=\left[\mathbf{I}_{MN} \ \text{vec}(\bar{\mathbf{H}}_k)\right]$. According to the analysis in~\cite[Theorem~1]{9183907}, we have  \begin{equation}
    \text{rank}(\mathbf{Y}_k^*)=\text{rank}(\Gamma\mathbf{I}_M-\mathbf{C}_k)\ge M-1, \forall k \in \mathcal{K}.
\end{equation}
Based on the Sylvester rank inequality, \eqref{C} indicates that
\begin{equation}
    \text{rank}(\mathbf{W}_k^*)\le M-\text{rank}(\mathbf{Y}_k^*)\le 1, \forall k \in \mathcal{K}.
\end{equation}}}

\bibliographystyle{IEEEtran}
\bibliography{mybib}

\begin{thebibliography}{10}
\providecommand{\url}[1]{#1}
\csname url@samestyle\endcsname
\providecommand{\newblock}{\relax}
\providecommand{\bibinfo}[2]{#2}
\providecommand{\BIBentrySTDinterwordspacing}{\spaceskip=0pt\relax}
\providecommand{\BIBentryALTinterwordstretchfactor}{4}
\providecommand{\BIBentryALTinterwordspacing}{\spaceskip=\fontdimen2\font plus
\BIBentryALTinterwordstretchfactor\fontdimen3\font minus
  \fontdimen4\font\relax}
\providecommand{\BIBforeignlanguage}[2]{{%
\expandafter\ifx\csname l@#1\endcsname\relax
\typeout{** WARNING: IEEEtran.bst: No hyphenation pattern has been}%
\typeout{** loaded for the language `#1'. Using the pattern for}%
\typeout{** the default language instead.}%
\else
\language=\csname l@#1\endcsname
\fi
#2}}
\providecommand{\BIBdecl}{\relax}
\BIBdecl

\bibitem{9424177}
Y.~Liu, X.~Liu, X.~Mu, T.~Hou, J.~Xu, M.~Di~Renzo, and N.~Al-Dhahir,
  ``Reconfigurable intelligent surfaces: Principles and opportunities,''
  \emph{IEEE Commun. Surv. Tutor.}, vol.~23, no.~3, pp. 1546--1577, 3rd Quart.
  2021.

\bibitem{9570143}
X.~Mu, Y.~Liu, L.~Guo, J.~Lin, and R.~Schober, ``Simultaneously transmitting
  and reflecting {(STAR)} {RIS} aided wireless communications,'' \emph{{IEEE}
  Trans. Wireless Commun.}, vol.~21, no.~5, pp. 3083--3098, May 2022.

\bibitem{10550177}
X.~Mu, J.~Xu, Z.~Wang, and N.~Al-Dhahir, ``Simultaneously transmitting and
  reflecting surfaces for ubiquitous next generation multiple access in 6{G}
  and beyond,'' \emph{Proc. of the {IEEE}}, vol. 112, no.~9, pp. 1346--1371,
  2024.

\bibitem{10740607}
H.~Li, Y.~Liu, X.~Mu, Y.~Chen, P.~Zhiwen, and X.~You, ``{STAR-RIS} in cognitive
  radio networks,'' \emph{{IEEE} Trans. Wireless Commun.}, vol.~23, no.~12, pp.
  19\,649--19\,666, 2024.

\bibitem{10439654}
J.-C. Chen, ``Designing {STAR-RIS}-assisted wireless systems with coupled and
  discrete phase shifts: A computationally efficient algorithm,'' \emph{{IEEE}
  Trans. Veh. Technol.}, vol.~73, no.~7, pp. 10\,772--10\,777, 2024.

\bibitem{10320337}
J.~Lei, T.~Zhang, X.~Mu, and Y.~Liu, ``{NOMA} for {STAR-RIS} assisted {UAV}
  networks,'' \emph{{IEEE} Trans. Commun.}, vol.~72, no.~3, pp. 1732--1745,
  2024.

\bibitem{8886335}
H.~Al-Obiedollah, K.~Cumanan, J.~Thiyagalingam, A.~G. Burr, Z.~Ding, and O.~A.
  Dobre, ``Sum rate fairness trade-off-based resource allocation technique for
  {MISO} {NOMA} systems,'' in \emph{Proc. IEEE WCNC}, 2019, pp. 1--6.

\bibitem{10264820}
A.~Papazafeiropoulos, P.~Kourtessis, and S.~Chatzinotas, ``Max-min {SINR}
  analysis of {STAR-RIS} assisted massive {MIMO} systems with hardware
  impairments,'' \emph{{IEEE} Trans. Wireless Commun.}, vol.~23, no.~5, pp.
  4255--4268, 2024.

\bibitem{10130543}
L.~Xue, K.~Wang, Z.~Yang, and M.~Peng, ``Max-min energy-efficiency fair
  optimization in {STAR-RIS} assisted communication system,'' \emph{IEEE
  Access}, vol.~11, pp. 51\,106--51\,116, 2023.

\bibitem{10311519}
Y.~Wang, Z.~Yang, J.~Cui, P.~Xu, G.~Chen, T.~Q.~S. Quek, and R.~Tafazolli,
  ``Optimizing the fairness of {STAR-RIS} and {NOMA} assisted integrated
  sensing and communication systems,'' \emph{{IEEE} Trans. Wireless Commun.},
  vol.~23, no.~6, pp. 5895--5907, June 2024.

\bibitem{9180053}
G.~Zhou, C.~Pan, H.~Ren, K.~Wang, and A.~Nallanathan, ``A framework of robust
  transmission design for {IRS}-aided {MISO} communications with imperfect
  cascaded channels,'' \emph{{IEEE} Trans. Signal Process.}, vol.~68, pp.
  5092--5106, 2020.

\bibitem{boyd2004convex}
S.~P. Boyd and L.~Vandenberghe, \emph{Convex optimization}.\hskip 1em plus
  0.5em minus 0.4em\relax Cambridge university press, 2004.

\bibitem{5447068}
Z.-Q. Luo, W.-K. Ma, A.~M.-C. So, Y.~Ye, and S.~Zhang, ``Semidefinite
  relaxation of quadratic optimization problems,'' \emph{IEEE Signal Process.
  Mag.}, vol.~27, no.~3, pp. 20--34, 2010.

\bibitem{9183907}
D.~Xu, X.~Yu, Y.~Sun, D.~W.~K. Ng, and R.~Schober, ``Resource allocation for
  {IRS}-assisted full-duplex cognitive radio systems,'' \emph{{IEEE} Trans.
  Commun.}, vol.~68, no.~12, pp. 7376--7394, Dec. 2020.

\end{thebibliography}

\end{document}